\documentclass[conference]{IEEEtran}
\usepackage[T1]{fontenc}
\usepackage[utf8]{inputenc}
\usepackage{graphicx}
\usepackage{tabularx}
\usepackage{balance}
\usepackage{pdflscape}
\usepackage{xspace}
\usepackage{url}
\usepackage{listings}

\usepackage{datatool}
\DTLloadrawdb[noheader,keys={b},headers={Class Name,\scriptsize{Diversity},\tiny{\# Methods}, \tiny{|TU|=1},\tiny{|TU|=2},\tiny{|TU|=3},\tiny{|TU|=4},\tiny{|TU|=5},\tiny{|TU|=6},\tiny{|TU|=7},\tiny{|TU|>7}}]{tab}{tab-amount-typeusages-table.csv}
\DTLremoverow{tab}{1}

\usepackage{numprint}

\usepackage{textpos}

\usepackage{framed}
\usepackage{mdframed}

\title{Empirical Evidence of Large-Scale Diversity in API Usage of Object-Oriented Software}
\author{Diego Mendez, Benoit Baudry, Martin Monperrus\\University of Lille \& Inria}
\date{}
  
\begin{document}
\begin{filecontents*}{tab-amount-typeusages-table.csv}
Name,Diversity,M6,1,2,3,4,5,6,7,>7
java.lang.String,2460,69,69,529,638,614,396,145,51,18
java.io.File,2166,47,45,373,775,613,264,69,17,10
java.lang.StringBuffer,1312,51,41,142,238,316,290,176,83,26
java.util.ArrayList,1236,36,36,179,307,328,236,115,29,6
java.lang.Class,872,62,62,333,286,115,45,18,8,5
java.util.List,724,31,30,149,235,194,86,23,5,2
java.lang.StringBuilder,643,44,42,92,139,142,132,63,22,11
org.eclipse.swt.widgets.Composite,639,227,135,222,131,86,41,16,4,4
javax.swing.JButton,625,143,83,119,141,102,72,43,21,44
javax.swing.JLabel,570,101,76,145,153,108,47,16,13,12
org.w3c.dom.Element,534,60,60,198,165,76,19,9,4,3
javax.swing.JPanel,530,108,77,115,116,112,65,31,12,2
org.w3c.dom.Node,516,39,38,128,150,95,46,29,18,12
java.util.HashMap,471,22,20,92,125,123,74,30,6,1
org.eclipse.core.resources.IFile,456,68,59,167,120,55,30,12,6,7
java.util.HashSet,453,23,23,75,134,120,77,19,5,0
org.eclipse.core.runtime.IPath,360,36,34,148,114,43,14,4,2,1
org.eclipse.swt.widgets.Label,312,97,56,83,68,68,23,8,3,3
javax.swing.JScrollPane,308,105,73,77,77,45,18,11,4,3
org.eclipse.swt.widgets.Display,247,157,108,86,34,9,4,1,2,3
org.w3c.dom.Document,209,61,56,79,45,17,6,3,2,1
org.eclipse.core.runtime.Path,192,48,25,61,62,33,7,4,0,0
org.eclipse.emf.common.util.EList,128,29,29,50,31,10,4,3,1,0
org.eclipse.core.runtime.IConfigurationElement,119,21,20,31,46,16,5,1,0,0
org.osgi.framework.Bundle,115,33,33,55,22,4,1,0,0,0
org.eclipse.core.runtime.IStatus,100,13,12,25,31,17,7,5,3,0
org.xml.sax.XMLReader,100,15,15,20,20,21,13,6,4,1
org.w3c.dom.Attr,94,22,19,33,22,14,3,1,1,1
org.eclipse.core.resources.IWorkspaceRoot,88,37,37,39,7,5,0,0,0,0
java.lang.Object,31,10,10,16,5,0,0,0,0,0
\end{filecontents*}

\newcommand\nbprojects{\numprint{3418}\xspace}
\newcommand\nbtotalusages{\numprint{9022262}\xspace}
\newcommand\nbtotalclasses{\numprint{382774}\xspace}
\newcommand\mtwoStringBuilder{\numprint{643}\xspace}
\newcommand\mtwoStringBuffer{\numprint{1312}\xspace}
\newcommand\mtwoString{\numprint{2460}\xspace}
\newcommand\moneString{\numprint{394959}\xspace}
\newcommand\msixString{\numprint{69}\xspace}
\newcommand\nbTypesLowDiversity{\numprint{382026}\xspace}
\newcommand\nbTypesHighDiversity{\numprint{748}\xspace}
\newcommand\nbTypesVeryHighDiversity{\numprint{48}\xspace}
\newcommand\mtwononjdk{\numprint{630}\xspace}

\maketitle

\begin{textblock*}{\textwidth}(0cm,-1.2cm)
\begin{center}
In Proceedings of the International Conference on Source Code Analysis and Manipulation (SCAM'2013)
\end{center}
\end{textblock*}

\begin{abstract}

In this paper, we study how object-oriented classes are used across thousands of software packages.
We concentrate on ``usage diversity'', defined as the different statically observable combinations of methods called on the same object.
We present empirical evidence that there is a significant usage diversity for many classes.
For instance, we observe in our dataset that Java's String is used in \mtwoString manners.
We discuss the reasons of this observed diversity and the consequences on software engineering knowledge and research.
\end{abstract}

\section{Introduction}

Gabel and Su \cite{gabel2010study} have published fascinating results, showing that most pieces of code of less than 35 tokens are redundant.
They appear elsewhere in the same project, or, for small sequences, elsewhere in the space of all ever-written software.
In ecology, a sister concept of redundancy is \emph{diversity}. In ecosystems, species are said to be redundant if they have the same functional role, and are said to be diverse if many different species occupy different niches.

There are many kinds of diversity in software \cite{Forrest:1997}.
In this paper, we focus on one kind of  diversity: the usage diversity of classes of object-oriented code.
Our main research question reads as follows.

\begin{framed}
Do all developers use a given class in the same way? or in diverse ways?'
\end{framed}

By ``usage diversity'', we mean ways of using a class in terms of method calls.
We consider software from the viewpoint of \emph{type-usages}, an abstraction introduced in \cite{Monperrus2010a,Monperrus2011b}.
This concept abstracts over tokens, control flow and variables interplay. 
In a nutshell, a type-usage is a set of method calls done on a variable, parameter or field in a code base. 
For instance, Figure \ref{fig:typeusage-example} presents a method body and three corresponding type-usages.

From a dataset of hundreds of thousands of Java classes, we have extracted millions of type-usages and measured their diversity (as defined by the number of different type-usages that can be observed).
For instance, we have found that the Java class ``String'' is used in \mtwoString different ways.
This is not an exception, our experiment provides us with empirical evidence that a large scale diversity exists in ``API usage''\footnote{We use the term ``API usage'' to reuse the same term  as close work \cite{LaemmelPS11}. In this case, ``API'' refers to ``Application Programming Interface'', which at the level of a class, is defined by the set of exposed methods (whether ``exposed'' means public, documented of callable).}.

\begin{framed}
We provide original observations  about the presence of diversity of API usage, founded on novel diversity measures about object-oriented code.  
\end{framed}

If the literature includes a large amount of work on the synthesis of artificial diversity in software systems \cite{Forrest:1997}, to our opinion, our work is the first study that empirically quantifies the presence of diversity in object-oriented code. Thus, an essential contribution of this paper is a set of new software metrics, inspired by biodiversity metrics, that quantify the amount and the structure of diversity of API usage.
Hence, our work can be classified as ecology-inspired software engineering research \cite{baudry2012,posnettdual}.

For most classes of our dataset, as expected, the API usage is limited to a handful number of ways of using the class.
However, we observe a large number of classes for which there are lots of different type-usages.
For us, as well as for many colleagues, this result is intriguing. It seems to contradict with known design principles that recommend to minimize the public interface and to strive for single responsibilities. 
The second half of the paper discusses those ``diverse classes'' (in particular the \nbTypesHighDiversity classes that have more than 100 type-usages). 

\begin{framed}
What are the factors causing such a high API usage diversity in object-oriented software? 

To what extent does this diversity question software engineering knowledge and research?
\end{framed}

We provide answers related to success of software libraries, API design, software repair and automated diversification.
We think that our pieces of evidence on API usage diversity shake up some established ideas on the nature of software and how to engineer it.
Some of our points are of speculative nature, but they aim at fostering a collaborative effort on understanding the factors behind this API usage diversity.
In particular, our long term goal is to translate the knowledge of API usage diversity into practice, by providing diversity-aware guidelines and tools to developers.

\begin{figure*}

\begin{minipage}{9cm}
\begin{framed}
\textbf {Source Code:}
\scriptsize
\begin{verbatim}

void saveNames(String inputPath) {
  ArrayList filenames = new ArrayList();
  File inputFile = new File(inputPath);
  if (inputFile.isDirectory()) {
    for (File f : inputFile.listFiles()) {
      filenames.add(f.getName()); }
  }
} 


\end{verbatim}
\end{framed}
\end{minipage}
\begin{minipage}{9cm}
\begin{framed}
\textbf {Abstraction:}
\scriptsize
\begin{verbatim}
// type usage #1 corresponds to "inputFile"
type:File 
calls:Constructor(String) isDirectory() listFiles()

// type usage #2  corresponds to "filenames"
type:ArrayList
calls:Constructor() add(String)

// type usage #3  corresponds to "f"
type:File
calls:getName()
\end{verbatim}
\end{framed}
\end{minipage}

  \caption{Illustration of the concept of ``types-usage''. An extractor transforms the Java source code at the left hand-side into the abstraction at the right hand-side. Type-usages abstract over tokens, control flow and variables interplay.}
  \label{fig:typeusage-example}
\end{figure*}

The rest of the paper reads as follows.
Section \ref{sec:background} gives some background on object-orientation and type-usages.
Section \ref{sec:studyapproach} describes our experimental design.
Section \ref{sec:diversity-first-results} exposes our empirical results and findings, while section \ref{sec:diversity_structure} investigates the deep structure of the observed diversity.
Section \ref{sec:discussion} discusses their implication.
Finally, related work (Section \ref{sec:rw}) and conclusion (Section \ref{sec:conclusion}) close the paper.

\section{Background}
\label{sec:background}
\subsection{Object-oriented software}
\label{sec:oo}

In object-oriented software, a class defines a set of functions (called methods) meant to be used in conjunction, in order to perform computations in a certain problem domain. For instance, in the problem domain of manipulating character strings, the Java class \verb|String| defines 76 methods to use and transform strings in a variety of manners. The term  ``object'' refers to an instance of a class.

In object-oriented software, variables can point to objects, and one ``calls'' methods on variables. Syntactically, this is written with a dot. Calling method "getFirstLetter" on a string variable is written \verb|a.getFirstLetter()|. The method operates on the data that is encapsulated within the object. Designing the scope of methods and where to put them is all the art of object-oriented design.

\subsection{Type-Usages}
\label{sec:tu}
We consider software from the viewpoint of \emph{type-usages}, an abstraction introduced in \cite{Monperrus2010a,Monperrus2011b}. A type-usage is a list of method calls on the same variable of a given type occurring somewhere within the context of a particular method body \cite{Monperrus2011b}.
Type-usages abstract over tokens, control flow and variables interplay. 

 An example is shown in Figure  \ref{fig:typeusage-example}. A call consists of the signature of the method to be called, that is, in Java, the method name, the parameter types, and the return type. 
Calls must be made on the same variable (whether local variable, method parameter or class field), are unordered (the location in source code is not taken into account) and unique (observing several times the same call on the same variable is not taken into account).
For instance, the methods \emph{void init(String)} and \emph{void init(File)} are considered as two different calls.
In the following, we will often refer to the type of a type usage as ``class''.

Importantly, many type-usages are of the same ``kind'' (same declared type, same set of calls).
In the following, when we use ``type-usage'', we mean this aggregated set of identical items. To refer to a concrete type-usage (say, the one corresponding to variable ``conn'' at line 318), we will use the term ``type-usage instance'' (programming terminology) or ``type-usage specimen'' (ecology terminology) .

\begin{table*}
\begin{tabularx}{\textwidth}{|p{4.3cm}X|}
\hline
\textbf{Abundance}&\\
$abundance_{project}(typeusage)$&  is the number of type-usages instances of a given type-usage for a single project  (in $[0, \infty[$).\\
$abundance_{ecosystem}(typeusage)$&  is the number of type-usages instances of a given type-usage in the ecosystem  (in $[0, \infty[$).\\
$abundance_{project}(class)$& is the sum of all type-usage instances that are typed by the same class in a given project ($\sum abundance_{project}(typeusage)$, in $[0, \infty[$).\\
$abundance_{ecosystem}(class)$&  is the sum of all type-usage instances that are typed by the same class in the ecosystem ($\sum  abundance_{project}(class) $, in $[0, \infty[$).\\
\hline

\textbf{Diversity}&\\
$diversity_{project}(class)$& is the number of different type-usages of a given class for a single project  (in $[0, \infty[$).\\
$diversity_{ecosystem}(class)$& is the number of different type-usages of a given class in the whole ecosystem   (in $[0, \infty[$). \\

\hline
\end{tabularx}

\caption{Ecology-inspired Diversity Metrics For Types-Usages.}
\label{tab:metrics}
\end{table*}

\section{Experimental Design} 
\label{sec:studyapproach}

Our experiment consists of collecting a large number of type-usages across open-source Java code.

\subsection{Dataset}

We have collected all Jar files present on a machine used for performing software mining experiments for 7 years.
A Jar file is an archive containing compiled Java code under the form of a collection of ``.class'' files.
We removed those duplicate Jar files which contain the same set of classes.
The resulting dataset contains \nbprojects Jar files.
The dataset only contains real code (mostly open-source code, but also binary proprietary code and student project code) and no artificial code that may have arisen along software mining.
It represents 11 GB of Java bytecode and refers to \nbtotalclasses different types (classes or interfaces).
The list of Jar files is given in the companion web page \cite{Mendez2013companion} and the raw data is available upon request.
In this paper, for the ecological metaphor, we call this dataset the ``ecosystem'' under study.

\subsection{Extraction Software}
The extraction software comes from our previous work \cite{Monperrus2011b}.
It extracts type-usages (described in \ref{sec:tu}) from Java code.
It uses the analysis library Soot \cite{vall99soot}.
It works at the method body scope for local variables and method parameters and class scope for method calls done on fields.
The extractor takes as input either Java source code or Java bytecode.

With Java source code, all dependencies must be known and present during analysis (i.e. all Jar libraries must be in the ``class path'').
With Java bytecode, thanks to Soot's ability to allow ``phantom'' references, the extractor can analyze Jars with unresolved dependencies.
According to our tests, having unresolved dependencies does not yield imprecision in the results. 
Whether mined from source code or bytecode, the extracted type-usages are mostly equivalent, since  the gap between Java source code and Java byte code is low, and quasi null at the abstraction level of type-usages. 
Since there are ``phantom'' references for most projects of our dataset, the experiment is based on bytecode analysis.
For sake of replication, the extraction software is available upon request.

\subsection{Metrics}
\label{sec:metrics}
The extraction of type-usages on our dataset yielded \nbtotalusages type-usage specimen.
We post-processed those type-usages to compute the metrics described in Table \ref{tab:metrics}.
There are two groups of metrics: \emph{``abundance metrics''} and \emph{``diversity metrics''}.
Metrics have two dimensions:
1) Whether they are computed at the type-usage or class level
2) Whether the are computed for a single project or for the whole dataset.

Those metrics are inspired from ecology.
The abundance of species is the number of specimen, we define the abundance at the level of type-usages and classes.
The abundance of a type-usage is the number of times it is observed in a given scope, i.e. the number of type-usage instances.

The richness of an ecosystem is one measure of diversity, it is the absolute number of species that can be observed in this ecosystem.
In our context, the richness of an object-oriented class is the absolute number of different type-usages found in a given domain. 
We call this metric $diversity_{ecosystem}(class)$.
A more precise definition is given in table \ref{tab:metrics}.

\section{Evidence of API Usage Diversity}
\label{sec:diversity-first-results}

For us, a very intriguing question is: what is the diversity of usages of object-oriented APIs? 
In other terms, do all developers use a given class in the same way? More formally, what are the values of $diversity_{ecosystem}$ as defined in table \ref{tab:metrics}?
For us, a class would be ``diverse'' if we observe many different type-usages of this type in the ecosystem under study.

\subsection{Abundance and Diversity Distribution}

\begin{table*}
  \centering\DTLdisplaydb{tab}
  \caption{The Diversity of 30 Widely Used API Classes and Their Number of Type-Usages per Size in Number of Method Calls.
The columns $|TU|=n$ give the number of type-usages consisting of $n$ method calls (e.g.; there are 69 type-usages of one single method calls for Java's String).}
  \label{tab:amount-typeusages-table}
\end{table*}

\begin{figure}
  \centering
  \includegraphics[width=\columnwidth]{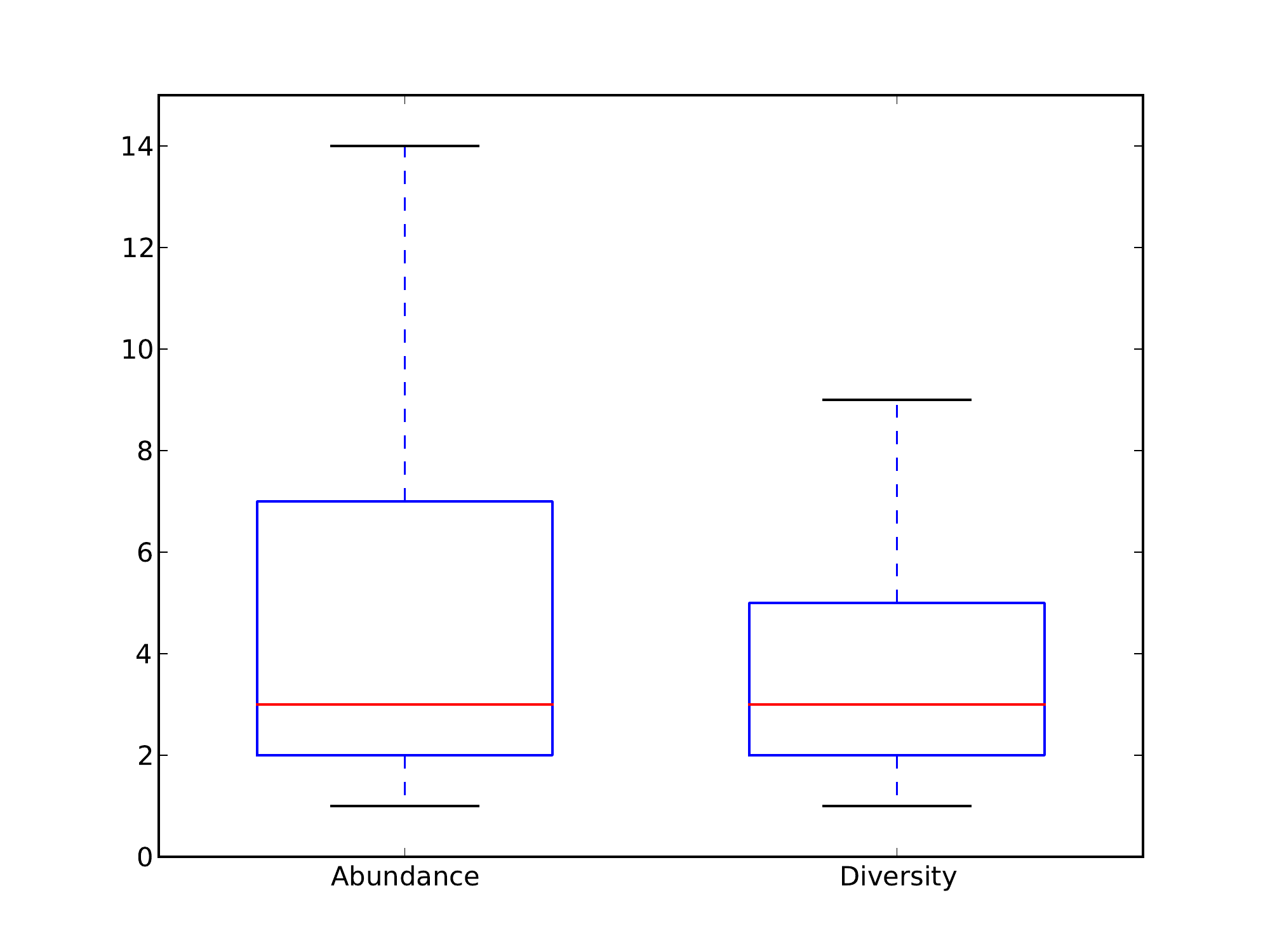}
  \caption{The Type-usage Abundance and Diversity of All Classes of the Dataset Under Study. The outliers are not represented for sake of scale.}
  \label{fig:boxplots-basic}
\end{figure}

Figure \ref{fig:boxplots-basic} shows the distribution of the abundance and diversity at the level of classes in the ecosystem as boxplots
($abundance_{ecosystem}(class)$ and $diversity_{ecosystem}(class)$ of Table \ref{tab:metrics}).
The median abundance is 4 (an abundance of 4 means that we have collected 4 type-usages for this class).
The abundance boxplot shows that across our \nbtotalclasses classes of our dataset, a large majority are used a small number of times.
This is due to the fact that many classes are only used in a single project (Jar file) of the dataset and within this project at most a handful of times.

The boxplot representing the distribution of diversity (second boxplot starting from left) shows that classes have a median number of 3 type-usages\footnote{Note that the maximum diversity of a class is necessarily its abundance in the case where each type-usage specimen is different. It thus makes sense that the median diversity is 3 given a median abundance of 4.}.
The upper quartile is 5. In other terms, for 75\% of the classes, we observe between 1 and 5 ways of using of the class.
\emph{However, the data contains many extreme points that are not represented on the boxplot since their order of magnitude dwarfes this low diversity.}

\subsection{Classes with High Usage Diversity}
Let us now concentrate on the upper quartile of the diversity metric, those classes with high usage diversity, 
In our dataset, there are \nbTypesHighDiversity classes for which we observe more than 100 different type-usages and \nbTypesVeryHighDiversity classes for which we observe more than 500 type-usages.
The extreme case is Java's String. For this class, we observe \mtwoString type-usages for (among \moneString type-usages specimen -- instances -- of type ``String'').

Table \ref{tab:amount-typeusages-table} gives the diversity of 30 diverse classes.
The first column is the diversity as defined in \ref{sec:metrics}. 
The second column is the number of called methods in the dataset.
The columns $|TU|=n$ give the number of type-usages consisting of $n$ method calls (e.g.; there are 69 type-usages of one single method calls for Java's String).
Those 30 classes come from the following stratified sampling: the 10 most used classes of the Java Development Kit (JDK) in number of projects, the 10 most used classes of Eclipse (an important sub-ecosystem of our ecosystem) and the 10 most used classes that are neither from Eclipse nor from the JDK. We refer to the latter as ``non-JDK classes'', we show them to show that usage diversity does not only appear in JDK classes. 
For instance, there are 534 different type-usages for W3's ``Element'' and 639 for Eclipse's Composite.

As programmers, we were really surprised by this richness.
Why were we surprised?
Probably because of the implicit principle of software engineering stating that an abstraction (whether function, class or method) should do one single thing (coined the ``Single Responsibility Principle'' by Robert Martin \cite{martin2003agile}).
In the perspective of type-usages, this principle reads as:
1) a class should have a small number of methods;
2) all methods should be used in the same way with some small variations.
However, in our opinion, having hundreds of type-usages for certain classes is not a small variation. 

Let us first deepen our understanding of this diversity before exploring the factors behind it.

\subsection{API Diversity Maps: A Graphical Visualization of OO Usage Diversity}
\label{sec:diversity-maps}

To help understand this diversity, we propose to represent the type-usages of a given class as a graph.
Each type-usage is a node in the graph.
The edges correspond to a subset relationship. If all the method calls of type-usage $x$ are contained into type-usage $y$, there is an arrow from $x$ to $y$.
The graph is laid out so that the largest type-usages are at the top and the smallest at the bottom.
We have seen that for certain classes, there may be hundreds of type-usages, hence hundreds of nodes in the graph, resulting in unreadable maps.
To overcome this issue, the map is parameterized with a threshold, responsible for filtering certain nodes.
The threshold filters $abundance_{ecosystem}(typeusage)$: if a type-usage has been observed in at least $N$ times, it is represented, otherwise it is discarded.
The rationale is that if a type-usage often appears, it is likely that the corresponding code has been written by many developers.
We call this visualization ``API diversity map''.

\begin{figure}
\includegraphics[width=\columnwidth]{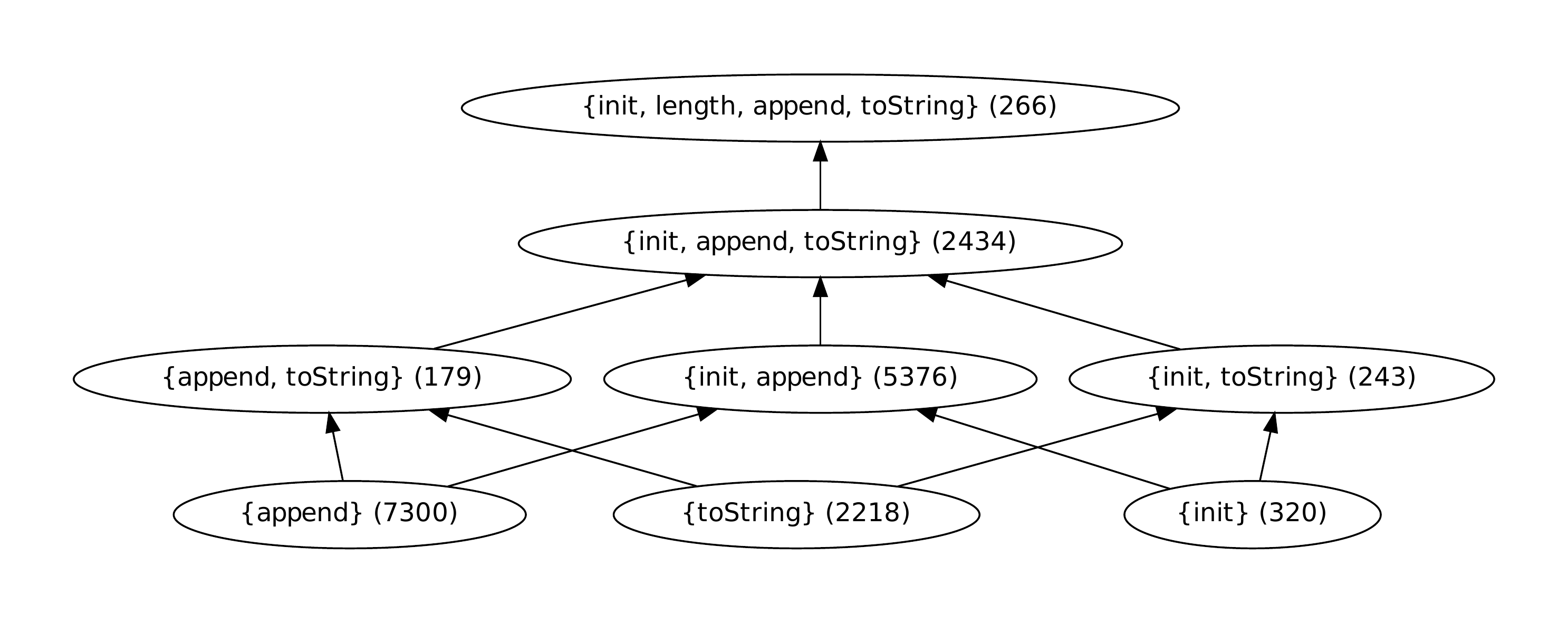}
   \caption{API Diversity Map of ``java.lang.StringBuilder''. The numbers in bracket is $abundance_{ecosystem}(typeusage)$.}
  \label{fig:api-map-Stringbuilder}
\end{figure}

Figure \ref{fig:api-map-Stringbuilder} gives the diversity map of Java's StringBuilder showing all type-usages that appear at least 150 times of the dataset.
The values for each type-usage correspond to $abundance_{ecosystem}(typeusage)$.
StringBuffer is a class used for manipulating strings in an efficient manner.
The map eventually contains 8 nodes which makes it very readable (in practice $diversity(StringBuilder)=\mtwoStringBuilder$ different type-usages).
This map is very layered, due to the semantics of edges (``subset of'').
One sees that there is a ``master'' type-usage in which all common methods of StringBuffer are used (``init'' refers to a constructor call).
One also sees that some type-usages are more popular than others.
For instance, \{init, append, toString\} appears 2434 in our dataset.
For developers who know StringBuilder, this reflects well its different usages.
For instance, on one end of the usage spectrum, one often only calls ``append'' on a StringBuilder passed as parameter.
On the other end of the usage spectrum, one uses all main methods of StringBuilder in a same method.

\begin{figure*}
\includegraphics[width=\textwidth]{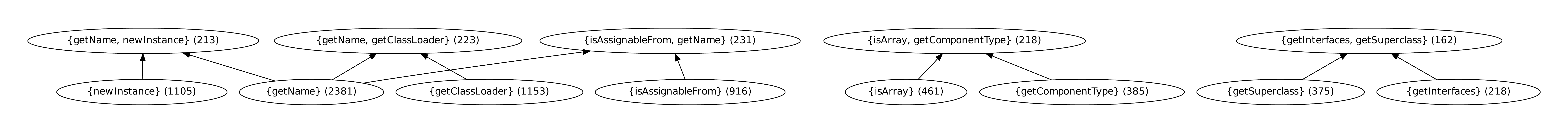}
   \caption{API Diversity Map of ``java.lang.Class''. The type-usage abstraction clearly captures different responsibilities.}
  \label{fig:api-map-Class}
\end{figure*}

Now consider the diversity map of Java's ``Class'' represented in Figure \ref{fig:api-map-Class}, the class handling the reflection of any object (the meta-object is obtained by calling ``getClass''). Compared to the diversity map of StringBuilder, we observe that: 
first the map is divided in three separated trees;
second, the top layer of the map is composed of 5 different type-usages.
Both phenomena  are due to the fact that Java's ``Class'' has different responsibilities:
creating objects (``newInstance''), proxying the current thread's class loader (``getClassLoader''), testing instance-of relationships (``isAssignableFrom''), handling Java array special semantics (``isArray''), and subtyping introspection (``getInterfaces, getSuperClass''). 
For this class, the visualization conveys in one glimpse that the class has different responsibilities.

\subsection{Why Is There Such a Large API Usage Diversity?}

Let us now discuss the reasons behind this API Usage diversity.

\subsubsection{An artifact in our extraction software?}
When we observed this phenomenon that has never been reported before the first thing we did was to check our extraction software.
We carefully browse the list of type-usages for classes (Map and String) to check whether 1) they make sense, 2) they actually appear in code.
The answer was positive. More generally, during our experiments, we have browsed many extracted type-usages and the corresponding source code for six months and this gives us confidence in our results.

\subsubsection{Type-usages Result From Combinations of Method Calls}
One reason behind this diversity is that type-usages are combinations of public methods.
The second column of Table \ref{tab:amount-typeusages-table} is the number of externally used methods on instances of those classes (in-class and inherited methods).
One sees that all diverse classes have a large number of methods, and that most methods appear in atomic type-usage with a single method call (e.g. for String, there are 69 used methods and 69 type-usages of size 1).
To check whether the usage diversity only depends on the number of methods for very diverse classes, we compute the Spearman correlation between the usage diversity and the number of public methods. The Spearman correlation is based on the ranks hence is independent of the exponential combinations of methods.
On the \nbTypesHighDiversity classes, the Spearman correlation is 0.25,
which is low.
The Spearman correlation is composed of numerical comparisons of the ranks of all pairs of classes.
A low value of 0.25 means that there are many pairs of diverse classes whose diversity and number of methods go in opposite directions.
Indeed there are 40\% of class pairs for which diversity goes in opposite directions (less methods but greater diversity).
This shows that the usage diversity is driven by more factors than only the number of public methods.

\subsubsection{Objects are Used across Different Methods}
Our analysis statically creates type-usages for local variables, method parameters and fields.
If at runtime, an object is passed from methods to other ones, our analysis would output several type-usages, while at the runtime object level, all method calls would be done on the same object.
For instance, let us consider a developer who wants to create a list, add elements and print them if the list is not empty. For some reasons, this developer would initialize the list in the class constructor, declare a new method for adding elements and a last one that prints the elements in a method that also  checks that the list is not empty. As a result, we would have 3 different type-usages: <init>, <add>, <isEmpty, get>.  We call those type-usages ``type-usage fragments''.
However, at the object level, the type-usage would be: <init, add, isEmpty, get>.
In the extreme case, if 10 methods are called in ten different methods, we would produce 10 type-usages, while there would be actually one. In such case, our diversity measures would be artificially 10x too big.

To explore this hypothesis, we propose to study the size of type usages of a given class. The idea is that if we only have very small type-usages, our static analysis has probably only captured small, non atomic type-usage fragments.
Let us consider again the diversity map of Figure \ref{fig:api-map-Stringbuilder}.
To some extent, the lower two layers of the diversity map correspond to fragments, because the corresponding objects necessarily all result from a call to the constructor.

Table \ref{tab:amount-typeusages-table} presents the distribution of type-usages per type-usage size for the 30 reference classes.
Recall that the columns $|TU|=n$ give the number of type-usages consisting of $n$ method calls.
Hence, the left-hand side columns contain small type-usages which are likely to be fragments. 
For instance, for Java's String (the first row), we observe in our dataset \msixString different type-usages of size 1.

So if one discards those small type-usages, do we still have a large diversity of type-usages?
The answer is yes. We observe many large type-usages, corresponding to method calls done on the same variable (and likely to the same object). Those type-usages are not artificial. 
\emph{Even with a conservative assumption that small type-usages are artificial fragments, we still observe a large diversity.}

% !TEX root = article.tex
% !TEX encoding = UTF-8
% !TEX spellcheck = en_US

\section{The Structure of Type-Usage Diversity} 
\label{sec:diversity_structure}
%% The Structure of Type-usage Diversity
%% The Type-Usage Dominance
%% API Entropy

We have observed in Section \ref{sec:diversity-first-results} that certain object-oriented classes give birth to a large diversity of type-usages.
Now we would like to understand the structure of this diversity: Are there type-usages that are much more used than the others?

Let us assume that we observe 1000 type-usage instances spread over 100 different type-usages.
If 800 of them are of the same type-usage, this would mean that the type-usage diversity is actually \emph{dominated} by a single one.
To reflect, we define the \emph{dominance} metric (called $dom$) as follows:

$freq_{ecosystem}(typeusage)$ is the frequency of a type-usage in the dataset (in $[0, 1]$).

$$
= \frac{abundance_{ecosystem}(typeusage) }{ \sum_i abundance_{ecosystem}(typeusage_i)}
$$

$dom_{ecosystem}(class)$ is the maximum observed frequency among type-usages referring to the same class (in $[0, 1]$). 

$$
dom_{ecosystem}(class) = max(\{freq_i) | type(i) = class\})
$$

\subsection{Type-usage Dominance}

\begin{figure}
  \includegraphics[width=\columnwidth]{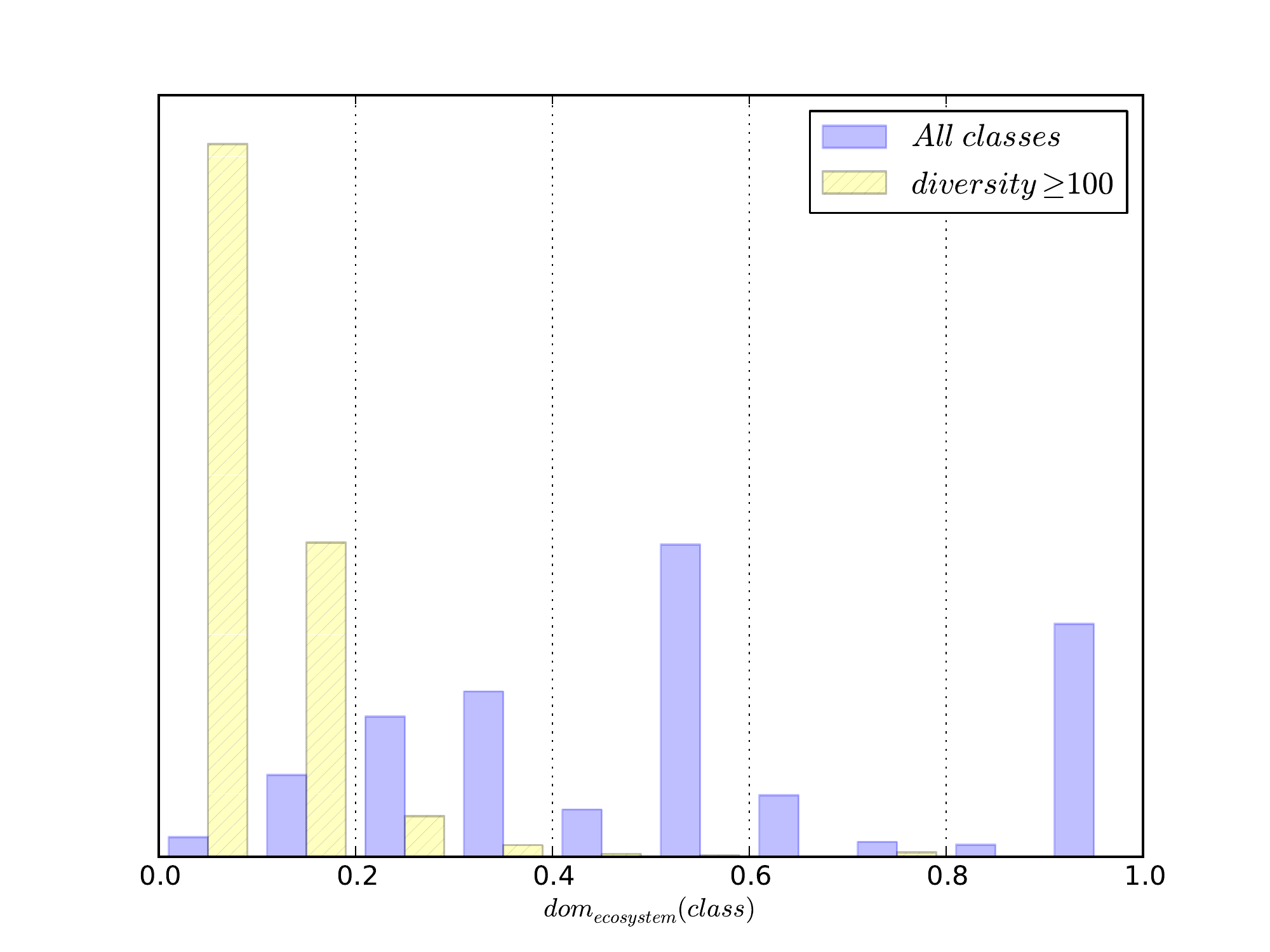}
  \caption{The Distribution of Dominance as an Histogram, for all classes of the ecosystem and for very diverse ones. Diverse classes have no dominant type-usages.}
  \label{fig:m3-ecosystem-histo}
\end{figure}

We have computed the type-usage dominance of the \nbtotalclasses classes of our dataset.
Figure \ref{fig:m3-ecosystem-histo} gives the distribution as an histogram (the plain, unhatched bars).
We observe two peaks around 0.5 and around 1.
A dominance of 1 means that all type-usage specimens of a given class correspond to the same type-usage, i.e. that there is no diversity at all. A dominance of 0.5 means that half of the type-usage specimens are identical.
Both cases are peculiarities of our dataset, corresponding to classes for which we observe one or two type-usage specimen.
The rest of the distribution contains ``dominated'' classes ($dom>0.5$) as well as classes for which there is no observed dominant type-usages (low dominance value, e.g. $dom>0.3$).
The latter correspond to classes where there is a real API usage diversity: nonetheless there are many type-usages but all of them are used in equal proportion.
Now, let us come back to to the high diversity observed for certain classes.

Let us concentrate on those \nbTypesHighDiversity classes for which we have observed more than 100 different type-usages. Are those classes really diverse?
Java's String has a dominance of 0.083, the most frequent type-usage is indeed not dominant.
Does this hold for the other very diverse classes as well?
The hatched bars of Figure \ref{fig:m3-ecosystem-histo} give the dominance distribution of those 750 very diverse classes.
Most classes have type-usage dominance lower than 0.2. The largest bin (the tallest hatched bar) corresponds to a dominance in the interval $[0, 0.1]$.
For those classes, there is no ``standard way'' of using the class and the type-usage diversity does not correspond to ``exotic variations''.

\begin{figure}
  \includegraphics[width=\columnwidth]{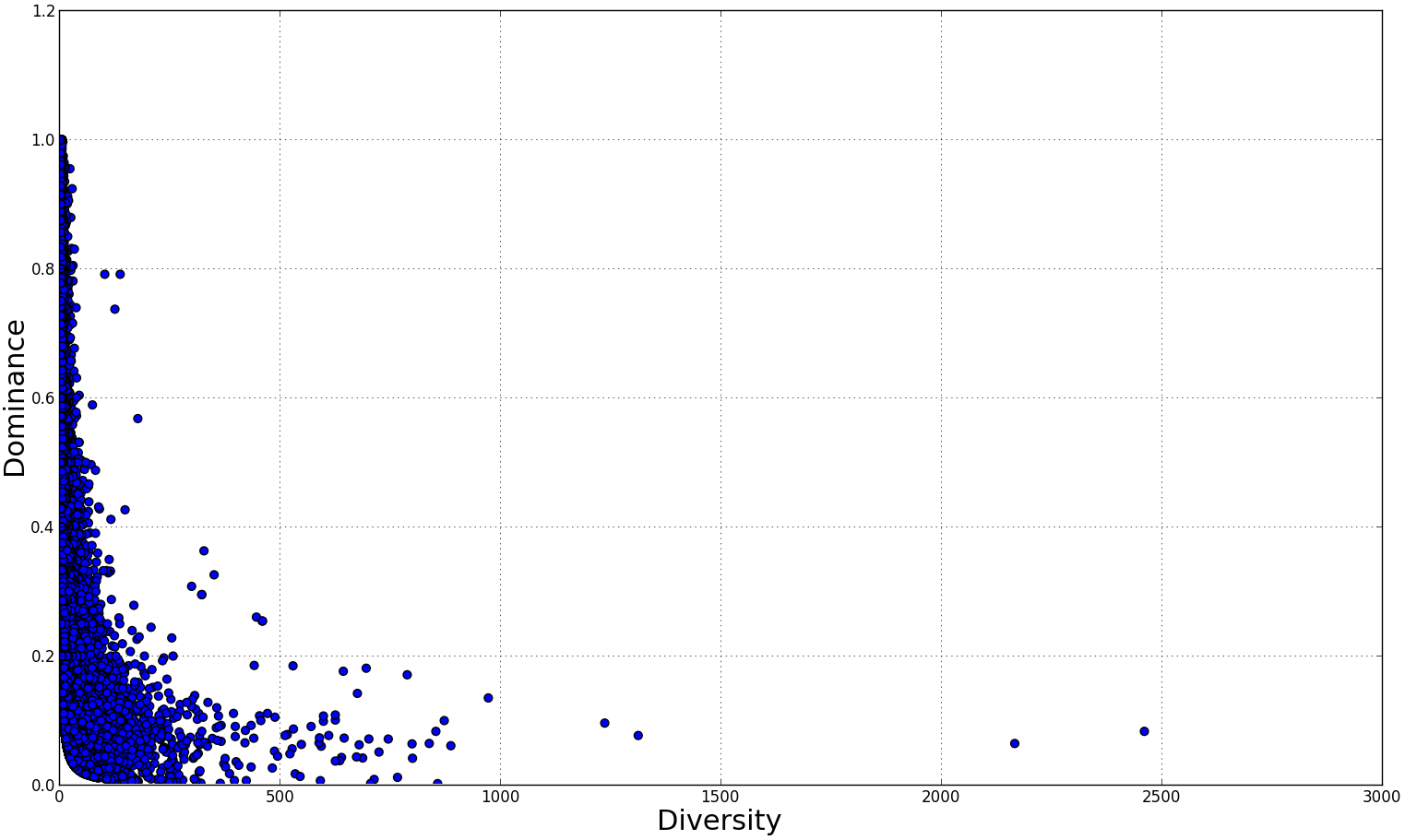}
 \caption{Correlation between Diversity and Dominance. Each point of the graphic is a class. The more diverse a class' type-usage, the less dominance.}
%TODO: graphics better legend and grid.
  \label{fig:m3-ecosystem-graphic}
\end{figure}

To further demonstrate this point, Figure \ref{fig:m3-ecosystem-graphic} plots the diversity and dominance values for each class of the ecosystem. The X axis is the diversity metric, the Y axis is the dominance metric. Each dot is a class. We can clearly see that there is a correlation between diversity and dominance: the more diversity, the less dominance. This confirms the findings on the \nbTypesHighDiversity most diverse classes.  Those pieces of evidence converge to state that \emph{the API usage diversity we have observed previously is actually a true diversity}.

%M2/M4: if M4 still increases, it means that there is a type-usage that includes even more usages. %Otherwise, there are new type-usages which are not a subset of it.

%There tends to be a similar relationship between M2/M3 and M2/M4 values. However, lately case is even more common and noticeable. When M2 increases, the value of M4 tends to decrease giving an exponential representation of the type among projects. This explainsa that when diversity into projects tends to increase and there does not exist an unique type-usage which represent a huge amount of other type-usages.
%On the other hand, for M2/M3 there are classes such as Java.util.HashMap where even with more than 60 different usages of a class, they represent more than 50\% or 60\% of the total amount of usages of the class.

\subsection{Usage Entropy of Classes}

The dominance metric reflects the skewness of the distribution of the abundance of type-usages. 
However, it neglects the distribution of the rest of the distribution, the $2^{nd}$ most abundant type-usage, the $3^{rd}$, etc.
To compute the overall skewness, we propose to use Shannon's entropy.
In ecology, Shannon's entropy is an established diversity metric \cite{good1953population} (``diversity index'' in the ecological terminology).
In our context, the entropy formula reads as follows:
 
$$
entropy(class) = - \sum freq(i) ln_2(freq(i) )
$$

where the $i$ are all observed type-usages of a class and $freq$ is an abbreviation of $freq_{ecosystem}(typeusage)$.
The entropy is correlated to diversity: the more entropy, the more diversity.

The entropy is maximum when all type-usages are equally distributed (i.e. of equal importance, with no dominance at all).
In this case, $ maxentropy(class) = - ln_2(diversity(class)) $.
This value is the theoretical maximum of the entropy, i.e. the maximum level of diversity.
For all classes of the ecosystem, let us draw $maxentropy(class)$ versus $entropy(class)$, in order to see whether the maximum diversity is often approached or not.

\begin{figure}
  \includegraphics[width=\columnwidth]{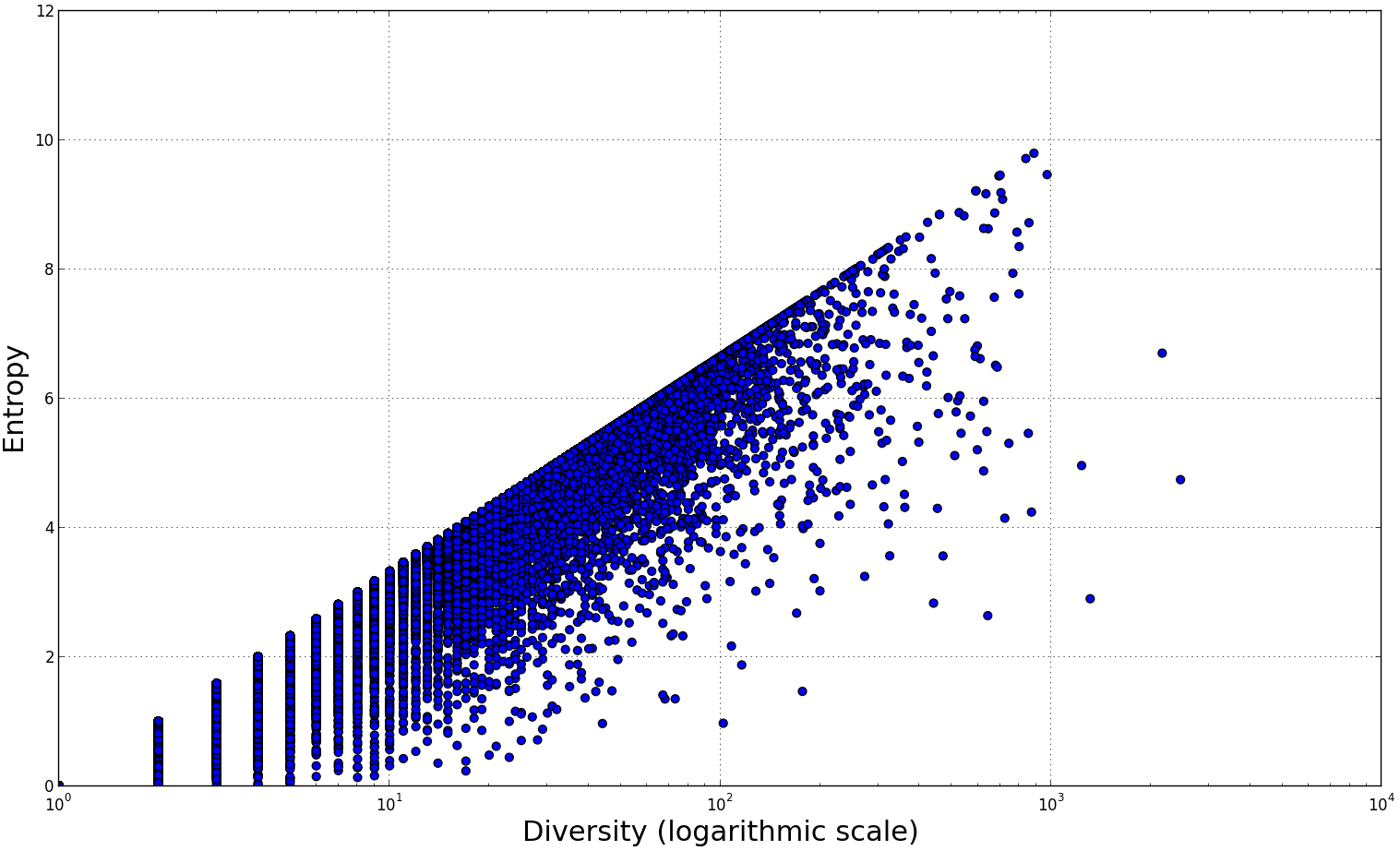}
 \caption{The Type-Usage Entropy of Classes (Y axis) as a function of the API Usage Diversity (X axis). Each point of the graphic is a class. Most classes are grouped just below the maximum entropy, i.e. the diversity is almost systematic.}
  \label{fig:entropy}
\end{figure}

Figure \ref{fig:entropy} is a scatter plot of the $entropy(class)$ (X axis on a logarithmic scale) versus $maxentropy(class)$  (Y axis).
Those axes represent the two components of what ecologists call ``species evenness''.
One dot is a class among the \nbtotalclasses classes of the ecosystem.
The diagonal lines emerging from the points correspond to the theoretical maximum entropy (when the type-usages are uniformly distributed).
There are no point for which $y>x$ for obvious theoretical reasons.
The vertical lines at the left-hand side of the figure correspond to all classes with a small number of type-usages (one line is $ln(diversity=3)$, one line is $ln(diversity=4)$, etc).
The main striking point of this figure is that \emph{the cloud of points sticks to the maximum entropy}.

First, it further validates the finding of Figure \ref{fig:m3-ecosystem-graphic}. While the dominance only takes into account the most frequent type-usages, the entropy reflects the skewness of the whole distribution. Since the points are grouped along the maximum entropy, with no gap between, this also shows there is a tendency to real diversity (the type-usages are all used frequently).
We would rephrase it as \emph{the API usage diversity is systematic}.

Second, let us concentrate on classes which have the same diversity value (according to metric $diversity$ of Table \ref{tab:metrics}). This corresponds to a vertical line of points. We see that those lines can be quite high, especially for low values of $diversity$. This means that there is a kind of a ``meta-diversity'': the distribution of type-usage abundance does not follow a simple rule for all classes.

\section{Discussion}
\label{sec:discussion}

We have observed a large-scale diversity in the usage of object-oriented classes.
Does this phenomenon impact our software engineering knowledge, beliefs?
Does it mean  something with respect to software engineering research and innovation?

\subsection{Diversity and Success}
\label{diversity-success}

Innovators try to write ``successful code''.
In a commercial perspective, to make a lot of money; in an open-source perspective , to gather a lot of users.
For an object-oriented library, ``successful'' means having many client pieces code.
For a class, ``successful'' means having many client type-usages across many different software projects.
Certain classes of the Java Development Kit are successful, as are classes of external libraries (e.g. the Apache Commons libraries).

How to write successful classes? There is no clear recipe and there are probably many factors influencing the success: technical, social and commercial.
However, it is generally accepted that a badly designed class has little chances to survive and become popular.

We have observed many classes that are successful (widely used across a large ecosystem), and that have a large number of public methods as well as a large diversity of possible different usages. Even if those characteristics are sometimes considered as bad design (as a violation of the single responsibility principle aforementioned), they do not prevent those classes to become successful. This holds for JDK classes as well as for non JDK classes (e.g. W3C's Node).
To sum up, according to our results, \emph{a high API usage diversity does not prevent success}.

We are also tempted to go further: \emph{if a class supports a high API usage diversity, it may favor its success}. The following section presents arguments in favor of diversity in API design.

\subsection{Diversity and Design}
Our results on API usage diversity raise many questions in terms of object-oriented API design.
We discuss in this question some of them. For educators who try to teach ``What good code is'', as we are, our results question certain conceptions and teach messages.

\paragraph{Diversity and Cognition}
When programming with object oriented APIs, the bulk of the cognitive load consists of remembering identifiers related to tasks (whether package, class or methods). With this respect, remembering one single class name is easier than remembering three of them. If Java's String would have been split in several classes, each one handling one fine-grain responsibility  (one subset of type-usages), this would have increased the cognitive load of developers. This argument applies to all classes and is related to research on API usability, in which we have not found studies about diversity. This argument would mean that, in terms of object-oriented API design, there is a trade-off between responsibility decomposition and usability. We think that future research on this point would be of great interest.

\paragraph{Diversity and Plasticity}
Second, let us define ``class plasticity'' as the ability of a class to be used in many different ways. Many factors influence the ``class plasticity''.
First, we have seen that the number of public methods increases the number of possible method call combinations, hence is correlated with the plasticity (although slightly as witnessed by the Spearman coefficient).
Second, all kinds of checks have an impact on the plasticity as well. For instance, overly restrictive pre-condition and post-condition checks hinder plasticity. 
We tend to think that a high usage diversity reflects a high class plasticity. 

\paragraph{Diversity and Reusability}
High usage diversity may correlate with reusability. It can reflect the fact that client code was able to use the class in ways that were unanticipated by the class designer. For instance, if one high level method is defined on three sub-routines, providing the subroutines as public would probably  provoke unanticipated reuse of those routines, which would consequently increase the class API usage diversity.
Having maps of API diversity as proposed in \ref{sec:diversity-maps} may guide reuse.
With those maps, developers are aware of whether certain type-usages are popular or not and can make informed decisions on how to use a class. Future research on how these api diversity maps impact a group of developers of different areas such as students, industry -related and research-related professionals would result of great interest to study.

\paragraph{Diversity and Immutability}
It is to be noted that one can add as many public methods to an immutable object without breaking anything: there are neither state-changing risks nor usage protocol issues. In other terms, an immutable class easily gives birth to a high API usage diversity.  Java's String being immutable, this argument probably contributes to the massive usage diversity we have observed.

\paragraph{Diversity and Testability}

Object-orientation has been a major concern in the software testing community: does it favor or hinder error finding?
In particular, increased encapsulation, modularity and coupling issues brought by the object-oriented paradigm led to a large amount of work that discuss the impact on testability \cite{bruntink2004, baudry2002, offutt2001}. Today, there is no doubt about the utility of object-orientation, and testers have found effective ways to reveal and fix errors in object-oriented code. However, the observations that we make in this paper seem to raise new questions about testability and maintainability of object-oriented libraries.  How to ensure that all possible type-usages are correct? Should there be one test per observed API usage (i.e. \mtwoString test cases for Java's String), or even one test per acceptable method call combinations?
This highlights a particularly intriguing relation between diversity and oracles, which we would put as diversity and correctness.
Does API usage diversity reflect a fuzzier notion of correctness? Does API usage diversity means that we can only have ``partial'' oracles?
This is an open question calling for future research on software testing.

\subsection{Diversity and Repair}

The type-usage abstraction has been introduced for sake of static bug detection \cite{Monperrus2010a,Monperrus2011b}.
In this previous research, our mantra was to find a definition of ``anomaly'' among type-usages, a definition that yields a low number of false positive.
An intuitive threshold on the abundance, even drastic, does not work.
However, we achieved a false positive ratio to the price of adding strong criteria in the definition of ``type-usage anomaly'': first, with respect to the context of the type-usage (the enclosing method), second, with respect to a type-usage distance expressed in terms of methods calls.
The new results presented in this paper illuminate our previous work: the diversity of type-usages makes it impossible to easily define an ``anomaly''.
When an observed world is too diverse, there is no such thing as ``anomaly'' or ``out of the norm''.
In general, we tend to think that the more diversity in code (resp. at runtime), the less possible it is to define high confidence static (resp. dynamic) bug detection rules. 

However, beyond bug detection, for automated bug repair, diversity may also as be a major opportunity.
The existence of a large number of similar, yet diverse type usages provides a wonderful `reservoir' of alternative code to fix bugs. This goes in the direction of recent results by Carzaniga and colleagues \cite{Carzaniga2013} showing that the API usage diversity and plasticity can be used to fix certain bugs at runtime.  In such cases, the diversity gives a kind of mutational robustness \cite{schulte2012software}.

\subsection{Diversity and Diversification}

In this work we make original observations about the presence of large scale diversity in software. This diversity is present and has emerged spontaneously through the development of a large number of Java classes. One question that emerges with the observation of this spontaneous emergence of diversity is: should we support or encourage the diversity in object-oriented software?
Beyond the impact of diversity on success discussed in \ref{diversity-success}, what about inventing techniques that automatically diversify a class API, using novel code synthesis mechanisms?

For example, let us imagine a developer who wants to use a class $X$. The developer calls a number of methods of this class' API, based on previous experiences with this API and a rather intuitive comprehension of what this class should do. There is a chance that the developer calls a method that is not part of the API, but that relates to the services offered by this API. If this case happens, there may be a possibility that the yet unknown method can be implemented in the as a combination of existing methods. One way to automatically diversify a class API would be to automatically synthesize this new method, using the code provided by the developer as the specification (if the code executes correctly, the generated method is correct). This kind of code synthesis would, by definition, increase the diversity of type usages over the API, and its principles would be similar to the theories underlying mediator synthesis for middleware interoperability \cite{blair2011, canal2008}.

\subsection{Recapitulation}

We think that our observations on object-oriented API usage diversity have questioned different parts of the software engineering knowledge in particular with respect to the principles of good API design. We also think that it opens new research questions in terms of API usability and software testing.

\section{Related Work}
\label{sec:rw}

Gabel and Su \cite{gabel2010study} have studied the uniqueness and redundancy of source at the level of tokens.
Our study explores a different facet: the diversity. 
In this paper, we have presented results at the level of object-oriented type usages, future work is needed to explore diversity at the level of tokens.

Baxter et al. \cite{Baxter2006} have studied the ``shape'' of Java software.
They discuss the empirical distribution  of many software metrics, in particular size based metrics. However, they don't discuss at all diversity metrics as we do in this 
paper.

At the level of object-oriented APIs, an early paper by Michail \cite{michail00} discusses object-oriented usage patterns that were observed in a large-scale study. He did not mention ``diversity'' although it was somehow implicit in the large reported number of patterns mined (51308 only for KDE classes). On the contrary, we focus on measuring, analyzing and understanding this diversity.

Ma and colleagues \cite{ma2006usage} only focus on Java classes and prevalence metrics. Laemmel et al. \cite{LaemmelPS11} talk about API footprint and coverage (the number of API classes and methods used within client projects). They do not mention the usage diversity.

To our knowledge, Veldhuizen \cite{veldhuizen2005software} is the only one who has looked at entropy in software in a similar meaning as we have. However, his point on entropy and reuse is more theoretical than empirical, and the presented results are at the level of low-level C library. To our knowledge, we are the first to report on the existence, with precise numbers, of large scale diversity at the API usage level.

Recently, Posnett et al. \cite{posnettdual} explored a facet of diversity in software development.
In their paper, they define the notions of ``artifact diversity'' and ``authorship diversity'' and extensively discuss the pros and cons of high diversity.
For instance; for a module, it is beneficial to have a high diversity of contributors.
Posnett et al. and we both specifically aim at measuring and understanding diversity in software.
But we focus on different facets: ``artifact diversity'' and ``authorship diversity'' are orthogonal to ``API usage diversity''.

\section{Conclusion}
\label{sec:conclusion}

We have mined \nbtotalusages type-usages which refer to \nbtotalclasses Java classes.
In this data, we wanted to specifically measure the \emph{diversity}, in the sense of ecological biodiversity.
To our surprise, we observed a large-scale usage diversity of API usage: \nbTypesHighDiversity classes are used in more than 100 different ways.
To our knowledge, this phenomenon has never been reported before.

We have started a discussion on the reasons and the impact of this observation on software engineering knowledge.
We look forward to gathering other diverse opinions to deepen the comprehension of this large-scale API usage diversity.
This paper reports on an empirical phenomenon, future work hopefully will find practical applications.
In particular, our long term goal is to translate the knowledge of API diversity into practice, by providing diversity-aware guidelines and tools to developers.

Also, it would be interesting to define measures of ``diversity'' at other levels of abstraction (e.g. tokens or control flow structures) to analyze the scale effect of this software metric \cite{posnettinference}.
Diversity may also vary depending on the application domains, and programming languages.
Furthermore, it is not clear how much the diverse type-usages of the same class are semantically different.
To conclude, the diversity advocated by Stephanie Forrest \cite{Forrest:1997} may have already emerged at many layers of the software stack and this work provides initial empirical insights about this phenomenon.

\section*{Acknowledgments}
This work is partially supported by the EU FP7-ICT-2011-9 No. 600654 DIVERSIFY project and the INRIA Internships program.
We thank Benoit Gauzens for detailed feedback.

\bibliographystyle{IEEEtran} 
\bibliography{biblio-paper.bib} 
\balance

\end{document}